# СЕГМЕНТНЫЙ ФОРМИРОВАТЕЛЬ ПУЧКА ЭЛЕКТРОМАГНИТНЫХ ВОЛН ДЛЯ ТЕРАГЕРЦОВОГО ЛАЗЕРА НА СВОБОДНЫХ ЭЛЕКТРОНАХ.

Г.Д. Богомолов[1], В.Д. Громов[2], В.В. Завьялов[1], А.А. Летунов[3].

[1]Институт физических проблем им. П. Л. Капицы РАН,
[2]Институт космических исследований РАН,
[3]Институт общей физики им. А. М. Прохорова РАН.

**Аннотация**

**Изготовлено и и исследовано сегментное зеркало для формирования зоны облучения мощным ТГц пучком облака пылевых частиц, моделирующих космическую пыль. Проведен численный расчет с учетом дифракционных эффектов и измерены основные характеристики зеркала в спектральном диапазоне генерации первой очереди ЛСЭ. Формирователь технологичен в изготовлении и может быть во многих случаях с успехом применен для концентрации мощного терагерцового излучения.**

## 1. Введение.

Для формирования пучка электромагнитных волн в системах генерации и регистрации излучения терагерцового[*] диапазона широко используются квазиоптические[**] [1, 2] отражательные элементы. Последние представляют собой обычно зеркала сложной формы, нередко составленные из отдельных элементов для упрощения технологии их изготовления.

В последнее время, в связи с распространением терагерцовых устройств за пределы узко научных применений, расширилось и применение квазиоптических устройств. Одновременно все более входят в практику мощные когерентные терагерцовые источники излучения, такие как лазеры на свободных электронах (ЛСЭ), примером которых служит Новосибирский ЛСЭ [3] Сибирского центра фотохимических исследований. Имеются особенности в использовании квазиоптических отражателей с когерентными источниками, которым уделено внимание в данном исследовании.

Разработка предназначена для экспериментов по взаимодействию терагерцового излучения с пылевыми частицами [4]. Исследование таких частиц, их взаимодействия с электромагнитным полем и между собой, - это одна из актуальных областей современной астрофизики [5], космологии [6], физики плазмы [7] и термоядерного синтеза [8]. В условиях разреженной космической плазмы, пылевые частицы существуют в виде рыхлых конгломератов наночастиц. Измерения сплошных (сплавленных или спрессованных) образцов неадекватно отражают физические, в том числе электромагнитные, свойства частиц, что привело к необходимости исследования пыли во взвешенном состоянии [9]. Задачей формирователя было создание пучка охватывающего объем со взвешенными частицами и при этом обеспечения необходимой интенсивности излучения в пучке, достаточной для разрушения слишком больших конгломератов, с одной стороны, и не слишком высокой, чтобы не вызвать эффекты электрического разряда (пробоя) в рабочем объеме.

Из-за высокой мощности ЛСЭ предпочтительными являются сплошные металлические отражатели ввиду их малого поглощения ТГц-излучения и высокой теплопроводности. Современные технологии позволяют изготовить металлическое зеркало сколь угодно сложной формы с точностью соответствующей длинам волн ТГц-диапазона в десятки и сотни микрон. Вместе с тем стоимость продуктов высоких технологий остается высокой, а к научным исследованиям все более выдвигаются требования сокращения затрат. С учетом этого представляют интерес зеркала простой формы, составленные из плоских сегментов, как рассмотреные в данной работе. Отчасти такой подход соответствует современным тенденции в оптике, которые привели, в частности, к тому, что лучшие в мире наземные телескопы имеют зеркала, составленные из относительно простых (сферических или асферических) элементов, называемых обычно сегментами [10].

Специфика формирования ТГц пучка новосибирского ЛСЭ связана с ситуацией, когда плотность мощности, имеющаяся в рабочих станциях недостаточна для проведения измерений, а использование обычного фокусирующего устройства с фокусным расстоянием, соответствующим размерам пространства вблизи рабочей станции ЛСЭ приводит к пробою в фокальном пятне. В данной работе рассматривается, довольно простое в изготовлении, устройство позволяющее сконцентрировать излучение в пятно необходимого диаметра, избегая при этом нежелательного перегрева или пробоя в элементах вблизи экспериментальной зоны.

## 2. Преимущества сегментного зеркала для формирования пучка ЛСЭ

Основное внимание в представленной работе было обращено на способы формирования пространственного распределения мощного ТГц пучка в зоне облучения пылевого облачка. Как следует из ранее сделанных оценок [4, 11], плотность потока ТГц мощности в пучке ЛСЭ в месте расположения рабочей станции может оказаться недостаточной для заметного нагрева взвешенной пыли. Поэтому необходимо было сконструировать и изготовить специальный формирователь пучка, исследовать его характеристики.

В соответствии с размерами экспериментальной установки, формирователь должен был концентрировать излучение в пятно диаметром около сантиметра на расстоянии всего полуметра от поверхности формирователя при максимальной плотности мощности в пятне на порядок меньше, чем была бы в центре фокального пятна обычного фокусирующего элемента с фокусным расстоянием 50 см. В условиях реальных ограничений на размеры устройства вакуумированный тракт должен быть достаточно компактным.

Использование обычных фокусирующих зеркал наталкивается на значительные трудности. Плотность мощности на окнах и других элементах тракта располагаемых вблизи пылевого облака должна быть как можно меньше (во всяком случае, не больше чем на входе формирователя), поскольку инициируемое пучком тепловое излучение этих элементов будет служить паразитным фоном при измерениях нагрева пыли. Получение адекватных задаче размеров перетяжки 1 - 2 см требует фокусных расстояний 10 - 20 м. А при фокусных расстояниях ~ 1 м из-за высокой пиковой мощности велика вероятность оптического пробоя в месте фокусировки в пылевом облаке.

Предложенный в работе [12] способ формирования пятна заданной конфигурации с использованием отражательных дифракционных оптических элементов (ДОЭ) не обеспечивает достаточной для наших измерений ахроматичности. Кроме того он сложен технологически и недоступен нам в изготовлении.

Поэтому была принята схема формирования перетяжки пучка вогнутой многогранной поверхностью, аппроксимирующей гладкую фокусирующую поверхность - мозаичным зеркалом из плоских фрагментов. При этом диаметр перетяжки ("фокуса") близок к характерному размеру граней, поэтому плотность мощности в фокусе можно достаточно просто регулировать размером граней, независимо от фокусного расстояния зеркала, что позволяет иметь дополнительные возможности для оптимизации длины оптического тракта облучения. В нашем случае формирование луча можно проводить мозаичным зеркалом на умеренных расстояниях до метра в фокусное пятно с характерными размерами порядка сантиметра.

По соображениям технологичности в качестве исходной гладкой поверхности выбрана часть поверхности внешней приэкваториальной области тора, рис.1. Здесь $R$ и $r$ большой и малый радиусы тора, а $\alpha$ угол падения. Параллельный пучок лучей, идущих в экваториальной плоскости тора (плоскости падения), фокусируется на расстоянии $0.5 \cdot (R+r) \cdot \cos\alpha$ от центра отражающего участка. А в случае пучка идущего в перпендикулярной плоскости фокусировка происходит на расстоянии $r/2\cos\alpha$. При равенстве этих величин мы получаем фокусировку цилиндрического пучка сравнимую по качеству с фокусировкой сферическим зеркалом при нормальном падении [13]. Мозаичная поверхность конструктивно подразделена на сегменты, вытянутые в направлении оси тора.

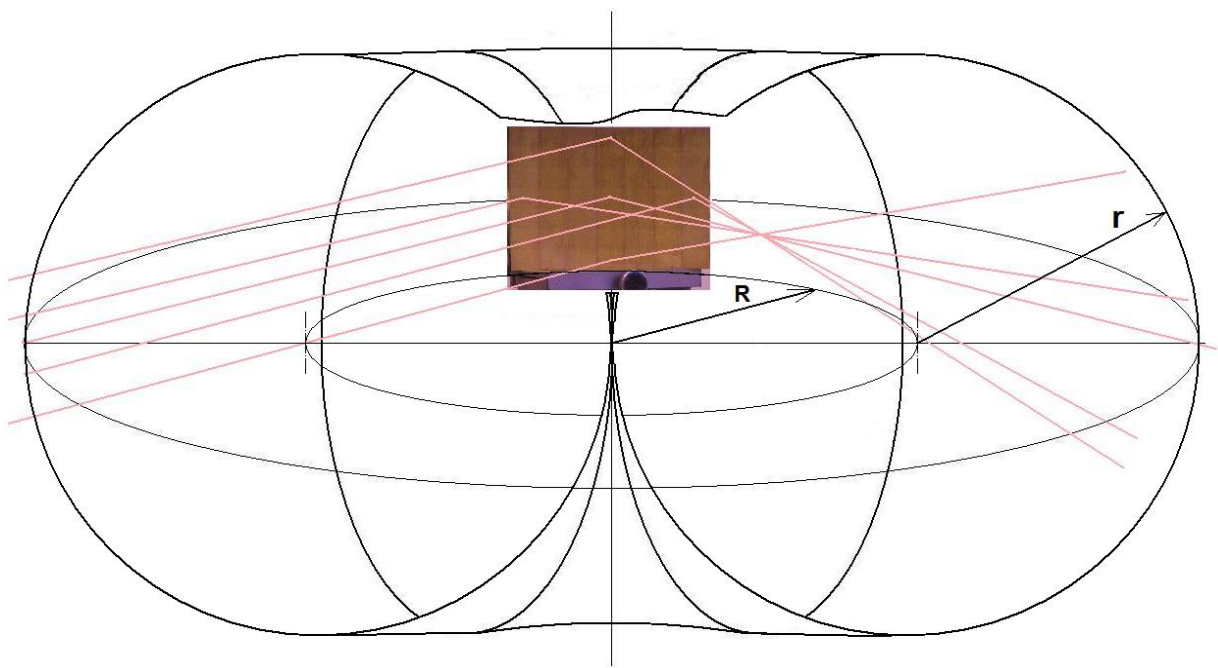

Рис.1. Тор со схематическим изображением мозаичного зеркала, с трапециевидными отражающими элементами, касающимися его поверхности центрами своих средних линий. Показан ход нескольких характерных лучей.

Ценой такого решения стал отказ от гладкости и монотонности радиального распределения пучка в перетяжке. В условиях многих экспериментальных задач такая неоднородность допустима. Необходимо так же отметить, что реальный профиль исходного пучка излучения ЛСЭ является достаточно сложным [12], что будет вести к некоторому уменьшению контраста результирующего распределения.

### 3. Конструкция.

Был разработан достаточно технологичный формирователь, состоящий из семи одинаковых сегментов (рис. 2), собираемых вместе на юстировочном устройстве, дающем возможность небольших перемещений каждого сегмента для составления из них единой заданной поверхности с достаточной точностью. Габаритные размеры рабочей поверхности ~140x200мм. Каждый сегмент является частью клина с углом при вершине 1° 38', и его рабочая поверхность состоит из 7-и трапециевидных площадок длиной 20мм, ширина прямоугольной средней площадки 28,2мм. В результате мозаичное зеркало состоит из 49-ти трапецеидальных плоских зеркальных элементов. Каждый из них касается в своем центре поверхности тора с большим и малым радиусами равными 70 см. Равенство радиусов обеспечивает одновременно с фокусировкой поворот луча на 90°, что необходимо в условиях эксперимента с подбрасываемой пылью на новосибирском ЛСЭ. При этом соотношении отверстие тора превращается в точку. Формирователь собирает пучок в перетяжку на расстоянии ≈ 0.5 м от вертикальной оси оптического канала станции ЛСЭ

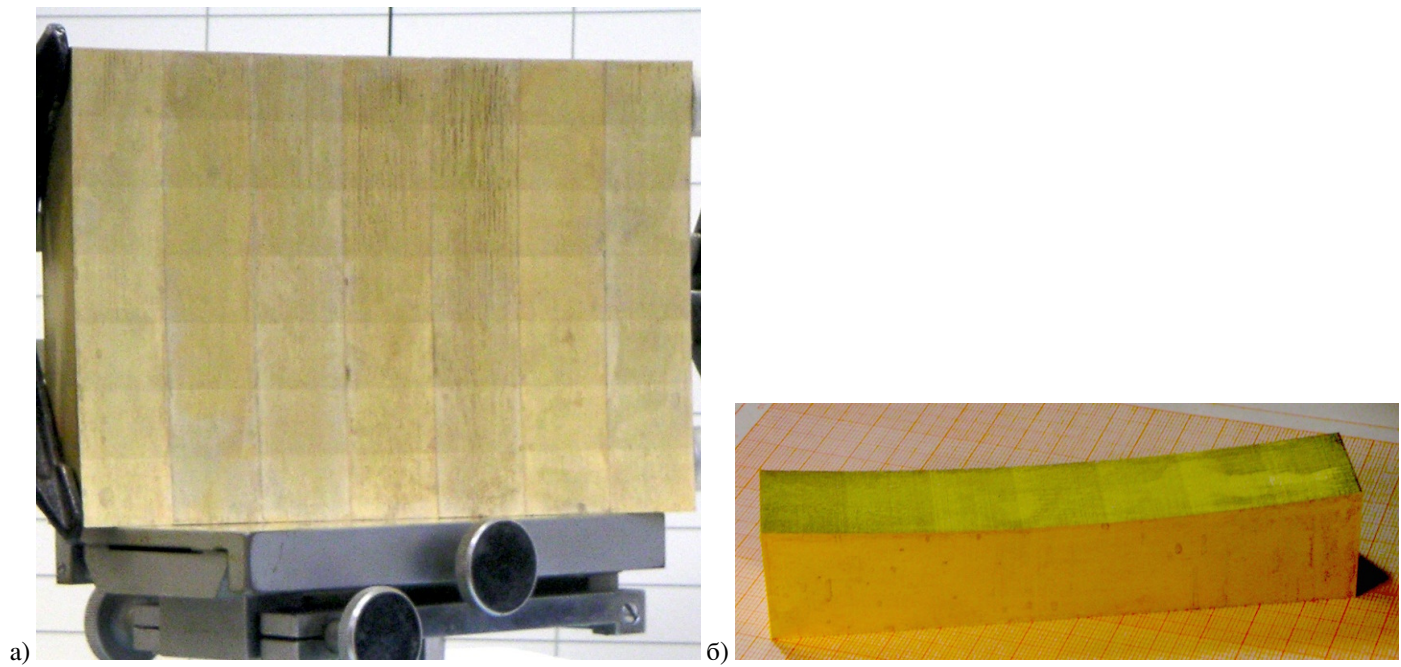

а)   б)

Рис. 2 а. Вид формирователя в сборе с юстировочными приспособлениями.
Рис. 2 б. Один из сегментов.

## 4. Расчет конфигурации "фокального" пятна

Хотя многоэлементные (мозаичные) зеркала находят все большее применение в практике, общие методы их расчета менее развиты по сравнению с оптическими системами составленных из единых сферических или асферических зеркальных поверхностей. Принципиальная разница - в эффектах интерференции между элементами волнового фронта, на которые разбивается фронт исходной волны при отражении от отдельных элементов зеркала. Мы провели расчет распределения интенсивности в фокальной плоскости двумя методами, сначала без учета интерференционных эффектов, а затем с их учетом. Как показано ниже, эти эффекты кардинально влияют на распределение интенсивности, что подтверждено и экспериментальными результатами, приведенными в следующем разделе.

Общепринятым универсальным методом расчета оптических систем при их конструировании является метод прогонки лучей (ray tracing). Применив его к мозаичному зеркалу можно оценить отклонения изображения от идеального в приближении геометрической оптики (очень коротких волн) [14]. При этом будет численно проверена наглядная модель, заложенная в основу конструкцию зеркала. Затем полученные отклонения изображения от идеального можно сравнить с оценками дифракционных эффектов и, тем самым в первом приближении выяснить, насколько конечность длины волны ( $\lambda > 0$) влияет на изображение исследуемой оптической системы.

Результаты "геометрооптического" расчета показаны на рис. 3, который демонстрирует распределение освещенности в "фокусе" мозаичного зеркала. Система координат при расчете выбрана так, что середина центрального сегмента зеркала была в начале координат $x=y=z=0$ под углом 45° к осям $x$ и $y$ перпендикулярно плоскости $xy$, так что исходные лучи параллельные оси $y$ отражаются в направлении оси $x$, а плоскость изображения есть множество точек $y, z$ с $x=F=50$ см. В проекция центрального сегмента на плоскость $yz$ имеет вид квадрата со стороной $p = 2$ см. Общее число сегментов равно $m^2$, где $m=7$.

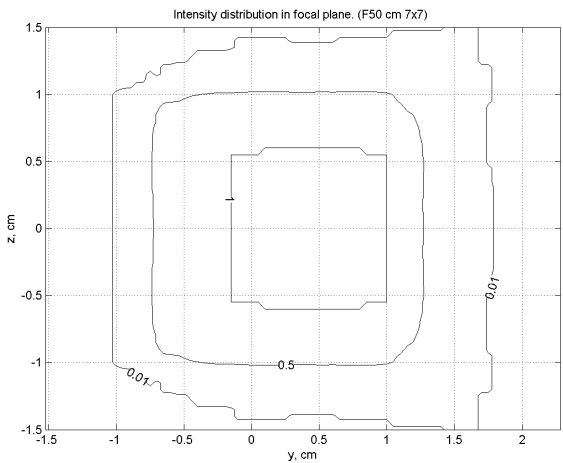 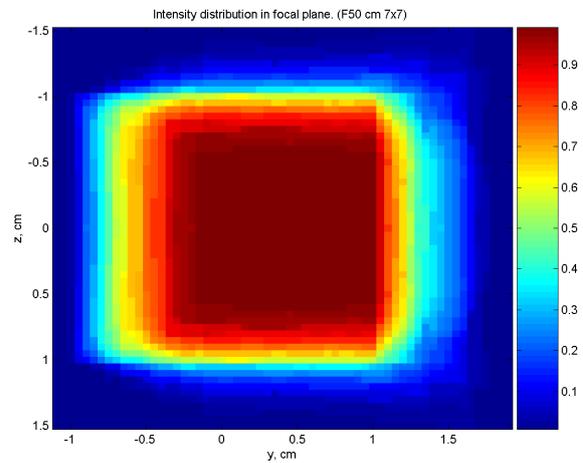

а) б)

Рис. 3. Распределение освещенности в плоскости перпендикулярной оси пучка.
а - показаны изофоты, б - то же изображение в искусственных цветах, отображающих интенсивность согласно шкале справа.

Зеркало было сконструировано так, чтобы каждый сегмент создавал пятно яркости в виде квадрата размером 2x2 см с центром вблизи начала координат на рис.3. Допустим некоторый сдвиг изображения, а размытость пятна не должна превышать дифракционное расплывание порядка $(\lambda F)^{1/2} = 1$ см. Из рисунков видно, что распределение интенсивности не слишком отличается от идеального изображения квадрата постоянной яркости. Контур пятна на уровне 1/2 (Рис. 3) отклоняется от сдвинутого "идеального" квадрата не более чем на 2 мм. В пределах этого контура интенсивность меняется не более чем в 2 раза, а в пределах центрального пятна размером около 1.2x1.2 см интенсивность постоянна. Таким образом задача решена для случая, когда можно пренебречь интерференцией сегментов, как в случае некогерентного, например теплового источника излучения, когда эффекты интерференции усредняются.

В случае когерентного источника, например лазера, эффекты интерференции создают зернистое изображение, состоящее из ярких точек, часто называемых спеклами (speckle). Они видны даже невооруженным глазом при рассеянии пучка оптического лазера. Эффекты интерференции и дифракции можно оценить, пользуясь принципом Гюйгенса-Френеля [14], который позволяет представить амплитуду волны в плоскости изображения $E(y_p,z_p)$, в приближении малых углов, как

$$E(y_p, z_p) = \int_S e^{ikr}/r \, E'(x,y,z) \, dS,$$

где волновое число $k=2\pi/\lambda$,
$r^2 = (x-F)^2 + (y-y_p)^2 + (z-z_p)^2$, $F$ - расстояние от центра зеркала до плоскости изображения, $x,y,z$ и $y_p,z_p$ - координаты на поверхности зеркала и в плоскости изображения, соответственно.
Распределение интенсивности в плоскости изображения $I = |E(x_p, y_p)|^2/4$.
Исходную плоскую волну можно представить как
$E'(x,y,z) = E_o(x,z) \, e^{-iky}$, где абсолютная величина амплитуды волны $E_o(x,z)$ - медленно меняющаяся функция $x$ и $z$. С учетом того, что величины $x, y, z, y_p, z_p$ много меньше $F$, можно учитывать отличие $r$ от $F$ только в показателе экспоненты, причем аппроксимировать $r$ как
$r \approx F-x + (x^2 + (y-y_p)^2 + (z-z_p)^2)/2F$.
В этом приближении численно оценивался интеграл

$$E(y_p,z_p) = F^{-1} \int\int_{-mp/2}^{mp/2} e^{ikr} E'(x,y,z) \, dx \, dz,$$

где $y$ задавался величинами $x, z$ и уравнением поверхности соответствующего сегмента зеркала.
Расчеты проводились для случая равномерной засветки $E_o(x,z)=\text{const}$ и для гауссова пучка

$$E_o(x,z) = \exp(-4(x^2+y^2)/D^2),$$

где $D = 12$ см. Картины распределения интенсивности имеют похожий вид, с той разницей, что для гауссова пучка основные интерференционные пики - шире, а вторичные пики - слабее. Результаты расчетов гауссова пучка представлены на рис. 4.

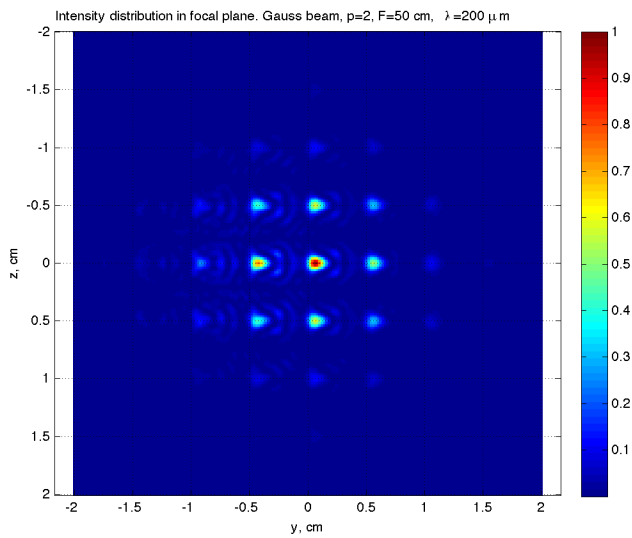 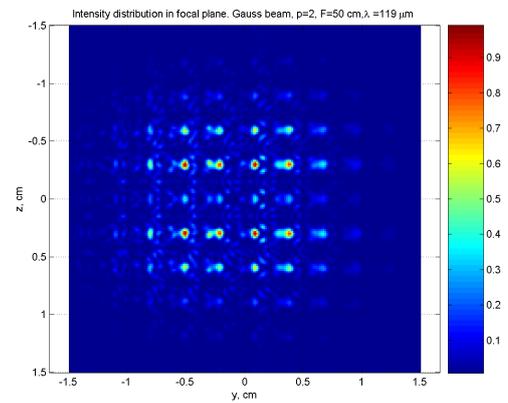

а) б)

Рис. 4. Расчетное распределение интенсивности, сформированное сегментным зеркалом из гауссова пучка.
Длина волны а) 200 мкм, б) 119 мкм.
Изображения имеют одинаковый масштаб, несмотря на различные размеры.

Следует отметить, что при заданных параметрах, дифракция плоской волны на элементе зеркала соответствует приближению Френеля. С другой стороны, по конструкции зеркала, расположение элементов таково, чтобы в большем масштабе сформировать фронт сферической волны, сходящейся к центру плоскости изображения. Как известно дифракция сферической волны формирует в фокальной плоскости картину дифракции Фраунгофера. В соответствии с этим интерференция пучков отраженных сегментами зеркала даёт картину напоминающую дифракцию Фраунгофера на экране с периодически расположенными отверстиями. При этом огибающая интерференционных пиков соответствует картине дифракции Френеля на отверстии с размерами единичного элемента.

Период основных пиков на расчетных распределениях совпадает с периодом пиков для дифракции Фраунгофера на перфорированном экране с тем же периодом расположения отверстий $p$.

$$P_x = P_y = \lambda F / p$$

Нетрудно заметить, что распределение интенсивности пиков соответствует картине дифракции одиночного сегмента (рис. 5). В частности оба распределения демонстрируют максимум в центре на 200 мкм, а на 119 мкм - минимум.

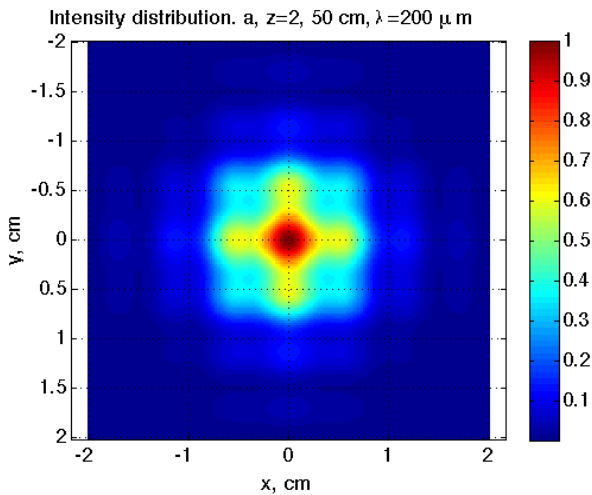 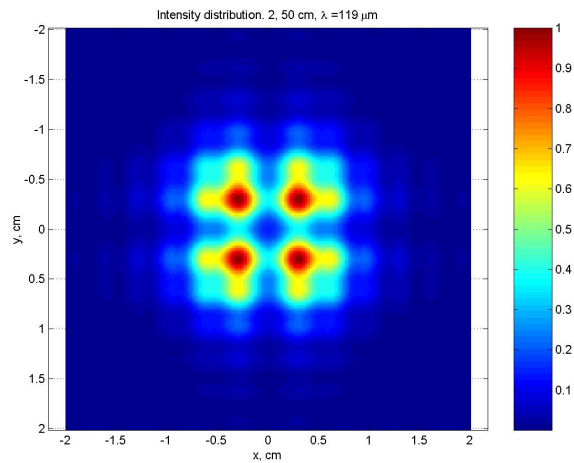

а) б)

Рис.5. Картина дифракции Френеля на квадратном отверстии со стороной $p = 2$ см.
Длина волны а) 200 мкм, б) 119 мкм.

Учет неравномерности поперечного распределения интенсивности гауссова пучка существенно не меняет этого результата. В основном гауссово распределение интенсивности пучка проявляется в подавлении слабых промежуточных пятен между основными точками концентрации энергии ЛСЭ. Учет немонохроматичности ЛСЭ в пределах 0.001-0.01, по оценкам также не существенно меняет картину. Таким образом, с учетом волновых эффектов, распределение интенсивности излучения в фокальном пятне сегментного зеркала имеет заметную неравномерность. Однако, пиковая интенсивность в «пятнах» на порядок ниже, чем при фокусировке вне осевым параболическим зеркалом. По оценкам, этого достаточно для того, чтобы избежать пробоя. Вместе с тем, для анализа экспериментальных данных необходимо достаточно хорошо представлять себе трехмерное распределение интенсивности в реальном лазерном пучке.

## 5. Изготовление формирователя

Был изготовлен рабочий образец зеркала из латунных сегментов. Сегменты были изготовлены на электроэрозионном станке с ЧПУ фирмы Фрязино-BEST. Линейные размеры были выдержаны с точностью 0,04мм, а угловые 0,002рад. Применение электроэрозионного станка обеспечило необходимую точность формы поверхности и достаточную чистоту обработки для работы в субмиллиметровом диапазоне. Качество полученной поверхности было проконтролировано на интерферометре белого света Zygo5000 (см. рис. 6).

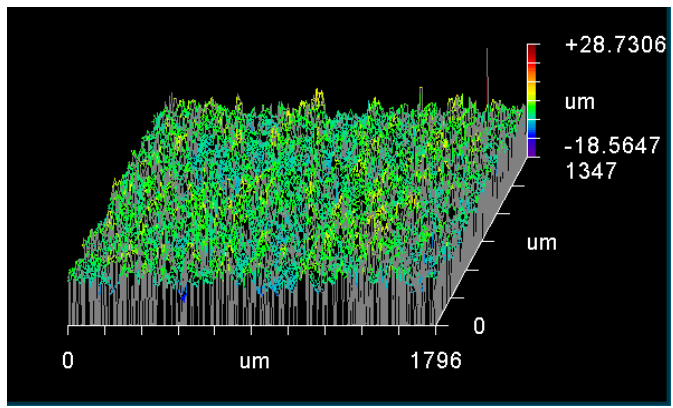

Рис. 6. Профилограмма участка сегмента, измеренная интерферометром Zygo5000.

Среднеквадратическое отклонение точек участка зеркала от плоскости - 3,5 мкм, что является очень хорошей величиной для диапазона длин волн 120-240 мкм.

Юстировка составного зеркала и первичный контроль геометрической точности изготовления и сборки проводился в оптическом диапазоне. Для этого использовалось излучение твердотельного непрерывного лазера с диодной накачкой и внутрирезонаторным преобразованием во вторую гармонику – DTL-318 компании "Лазер-компакт". Зеленое излучение с длиной волны 532 нм, мощностью 200 мВт, разведенное телескопом на основе коллиматора оптической скамьи ОСК-2 до рабочего светового диаметра отражателя – 140мм. Для получения оптического отражения от шероховатой в оптическом диапазоне поверхности отражателя на рабочие поверхности сегментов временно наклеивалась лавсановая пленка. Эта пленка прозрачна и в субмиллиметровом диапазоне, что позволяло проводить первичную настройку схемы измерений в субмиллиметровом диапазоне одновременно с оптической юстировкой зеркал.

Выставив предварительно угол падения, и контролируя подвижным белым экраном поперечное распределение освещенности, достаточно легко удается свести все отражения в видимом свете в требующийся яркий квадрат ~ 20x20 мм.

## 6. Измерение характеристик отражателя в терагерцовом диапазоне

Для изучения функционирования отражателя в спектральном диапазоне излучения ЛСЭ были проведены модельные измерения с одномодовым H2O лазером [15], работавшим на длинах волн 118.6 и 220 мкм. Наиболее детальные измерения проведены на 118.6 мкм.

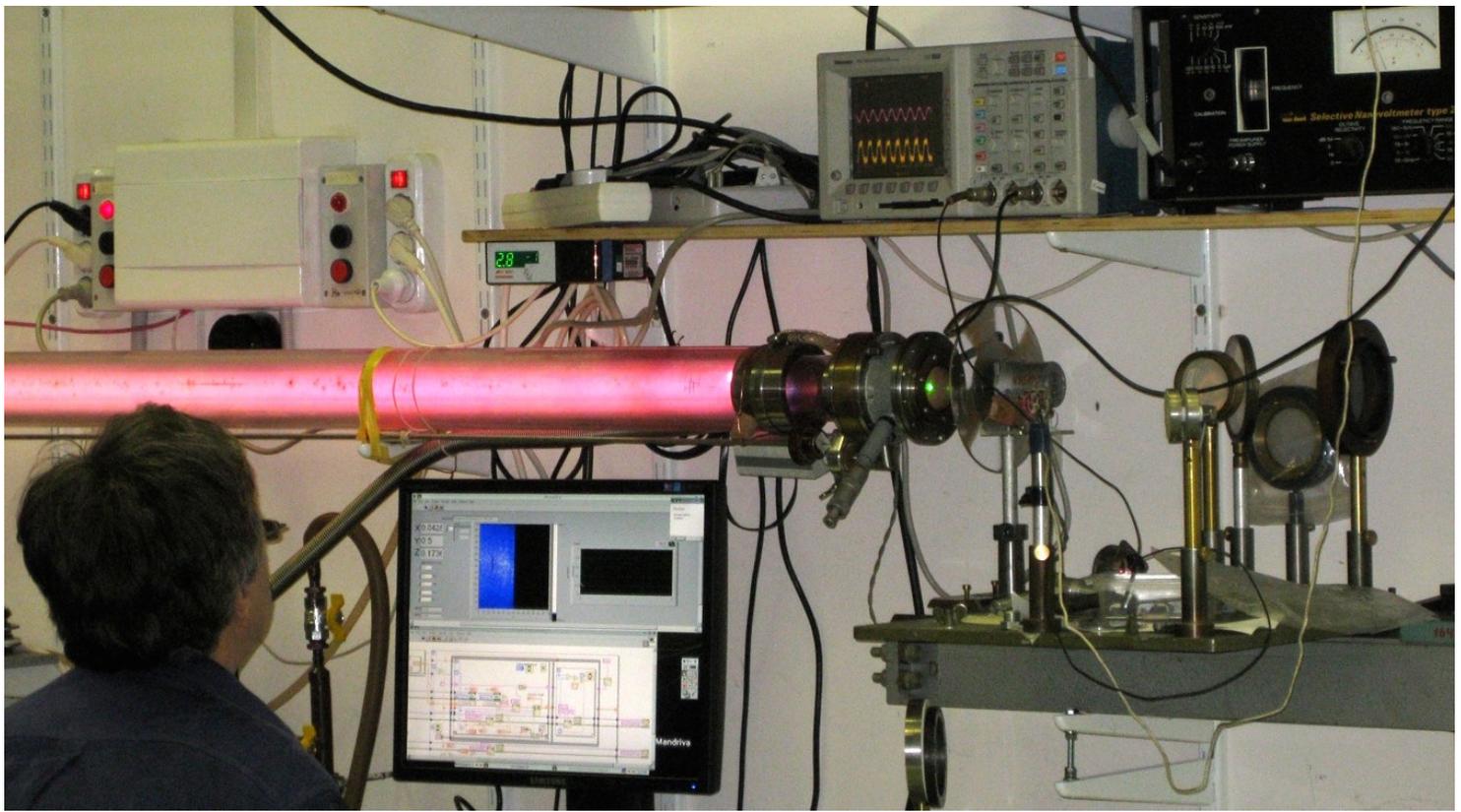

Рис. 7. Общий вид измерительной установки с H$_2$O-лазером.

Полное заполнение апертуры сегментного зеркала обеспечивалось за счет естественной расходимости гауссова лазерного пучка. Для этого от выходного окна лазера до сегментного зеркала луч проходил, отражаясь от плоского поворотного зеркала, оптический путь длиной около 12-ти метров. Для обеих линий выходная мощность лазерного пучка была порядка 2 - 5 мВт, что примерно на пять порядков меньше средней мощности ЛСЭ. В этих условиях только за счет очень высокой чувствительности детекторов можно достичь хорошего отношения «сигнал-шум» при измерениях с достаточно высоким пространственным разрешением.

С этой целью в качестве детекторов были использованы охлаждаемые жидким гелием германиевые кристаллы (Ge:B) - для приема излучения с длиной волны 118.6 мкм и кристаллы InSb в резонансном магнитном поле - для 220 мкм. Приемник с двумя детекторами, размещенными в общей металлической полости в одной вставке, помещался в гелиевый дьюар[16]. Излучение из области перетяжки луча, сформированной сегментным зеркалом, подводилось к детекторам по системе лучеводов в виде состыкованных шарнирами с зеркальными элементами отрезков тонких телескопических трубок из нержавеющей стали с внутренними диаметрами ~ 3-4мм. Пространственное разрешение определялось отверстием в металлической вставке на входном торце лучевода. Выходная трубка лучеводной сборки была жестко скреплена с входом лучевода низкотемпературного датчика.

Область перетяжки в «фокальной» плоскости зеркала размером ~ 40x40 мм зондировалась контролируемым перемещением входного торца лучевода. Сканирование реализовывалось за счет электромеханической системы, поставленного вертикально на переднюю торцевую стенку двухкоординатного самописца BRIANS. В его каретке был шарнирно закреплен входной торец лучевода, а на входные усилители подавались меандры от двух цифровых, управляемых ПК генераторов Agilent.

Сигналы криогенного детектора и опорного пироэлектрического датчика подавались на селективные вольтметры и в конечном итоге так же регистрировались в ПК. Измерения были полностью автоматизированы с использованием софта LabView. Для качественной записи с достаточным пространственным разрешением потребовалось использовать большие времена накопления, и как следствие медленное сканирование. На рис. 8 представлено распределение интенсивности, снятое на длине волны 118.6 мкм за 4 часа.

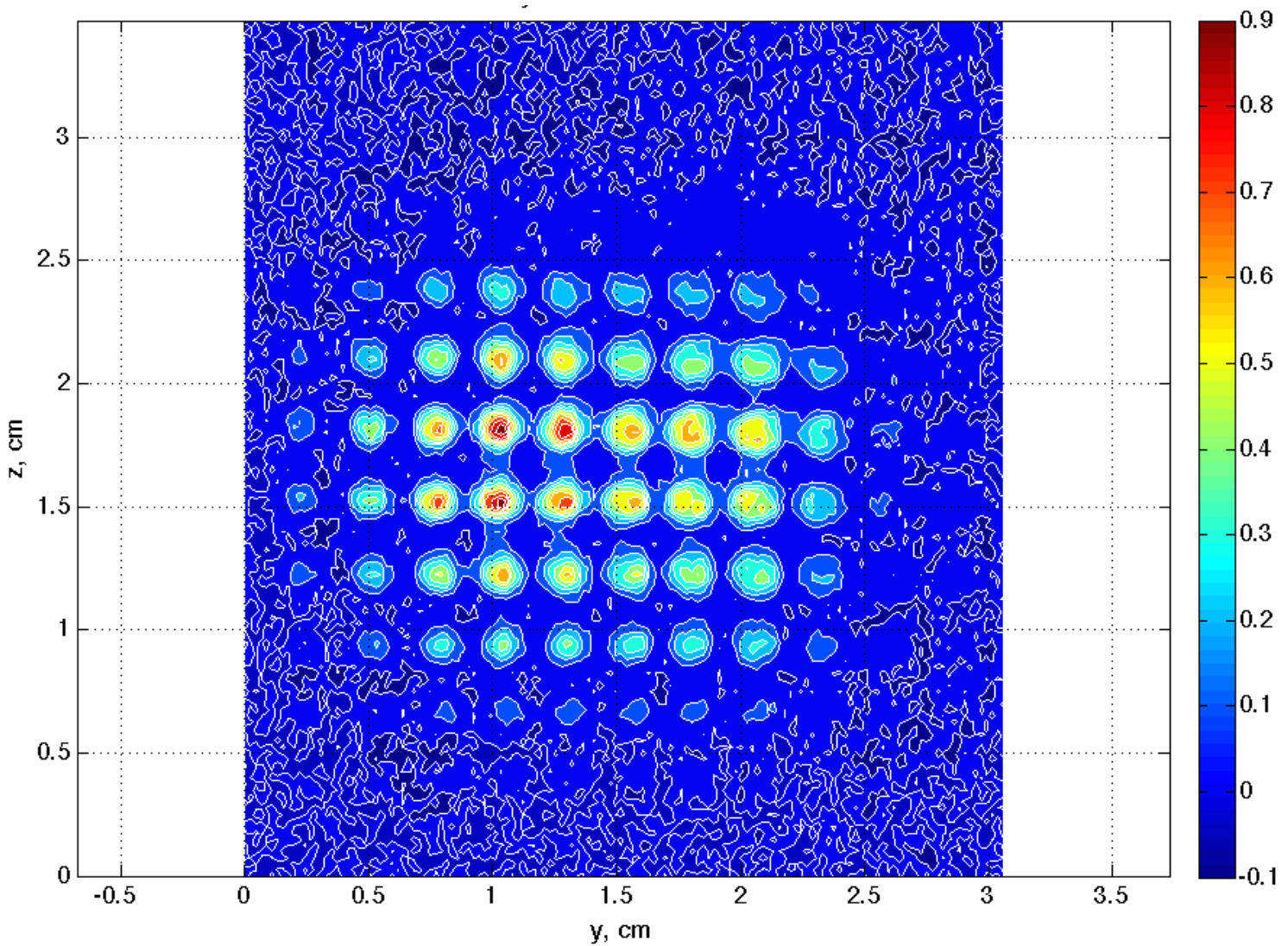

Рис. 8. Результаты измерений распределение интенсивности изображения, сформированного сегментным зеркалом. Источник излучения - газоразрядный лазер на парах воды. Длина волны 118.6 мкм.

Максимум интенсивности на рисунке равен единице, контуры интенсивности проведены через 0.1. На цветовой шкале указано нижнее значение интервала интенсивности. Так интервал -0.1 — 0 обозначен темно-синим цветом как -0.1. Естественно, отрицательные значения на картинке являются проявлением низких положительных значений интенсивности в следствие шумов измерений, которые и создают мелкие детали на периферии картинки.

Было получено несколько аналогичных изображений на 118.6 и 220 мкм. К сожалению изображения на 220 мкм имели существенно худший сигнал/шум, поэтому для них можно было сказать о качественном согласии измерений с расчетом, а в численном анализе пришлось ограничится данными 118.6 мкм. В целом, измеренные распределения имеет ту же структуру, что расчетные. Пятно в фокальной плоскости состоит из узких и ярких интерференционных пиков.

Расстояния между пиками не являются таким постоянными, как в расчете. Оно находилось в интервале 2-4 мм, но его среднее значение составило величину около 3 мм, в соответствии с выше приведенным соотношением $P_x = P_y = \lambda F/p$. По-видимому, отклонения связаны со случайным характером ошибок юстировки сегментов зеркала, тогда как при расчете отклонения от идеальной формы носят регулярный характер.

Ширина пиков в этом и последующем измерении составляла от 1 до 2 мм, тогда как в расчетах для той же волны она была 0.7 ±0.05 мм. При этом повторные измерения были проведены с лучшим пространственным разрешение, что, однако, не существенно сказалось на ширинах пиков, но заметно ухудшило отношение сигнал/шум. Отличие ширин пиков нельзя отнести к характеристикам зеркала, но можно объяснить вдвое меньшим эквивалентным диаметром пучка лазера по сравнению, поперечными размерами пучка излучения ЛСЭ, одинаково использованными при расчетах на обеих длинах волн.

Диаметр фокального пятна оценивался по огибающей главных пиков интенсивности на уровне половины интенсивности наиболее яркого пика. В расчетном распределении он составлял 13 мм, а в измеренных распределениях оценки находились в пределах 12-18 мм и, таким образом, и по этому параметру не противоречат расчетам.

## 7. Обсуждение результатов

В отличие от расчетного распределения (рис. 4б), на измеренном распределении (рис. 8) не виден центральный минимум интенсивности, свойственный на этой волне дифракции от одного сегмента (рис. 5.б). Это не удивительно, если пики интенсивности расположены так, что не совпадают с минимумом. Несовпадение есть уже на рис. 4б по оси *y*, где центральные пики имеют *z*=0, но *y*≠0. Минимум не проявится, если в координатах, привязанных к центру зеркала нет пиков с *z*=0 или *x*=0.

Такой сдвиг возможен, если симметрия по оси *z* нарушена неточной юстировкой сегментов или входного пучка.

Практической необходимости разрабатывать методы прецизионной юстировки нет. Достаточно при использовании данного зеркала (как формирователя пучка излучения) принимать во внимание высокую чувствительность детальной картины расположения пиков от юстировки. Это объясняет то, что в повторных измерениях распределения интенсивности, с небольшими подстройками юстировочных элементов, наблюдалось заметное изменение положения деталей, при сохранении общих параметров распределения. Приведенные в разд. 4 расчеты соответствуют области, где пики интенсивности выражены наиболее контрастно. Аналогичные расчеты с удалением от фокуса хотя бы на 1-2 см демонстрируют переход к более хаотичной спекл-картине и, тем самым, вопрос с неравномерностью распределения интенсивности в пятне становится не столь существенным.

Можно сравнить с теорией экспериментальные распределения вероятности значений интенсивности, полученные статистической обработкой изображений (рис. 9).

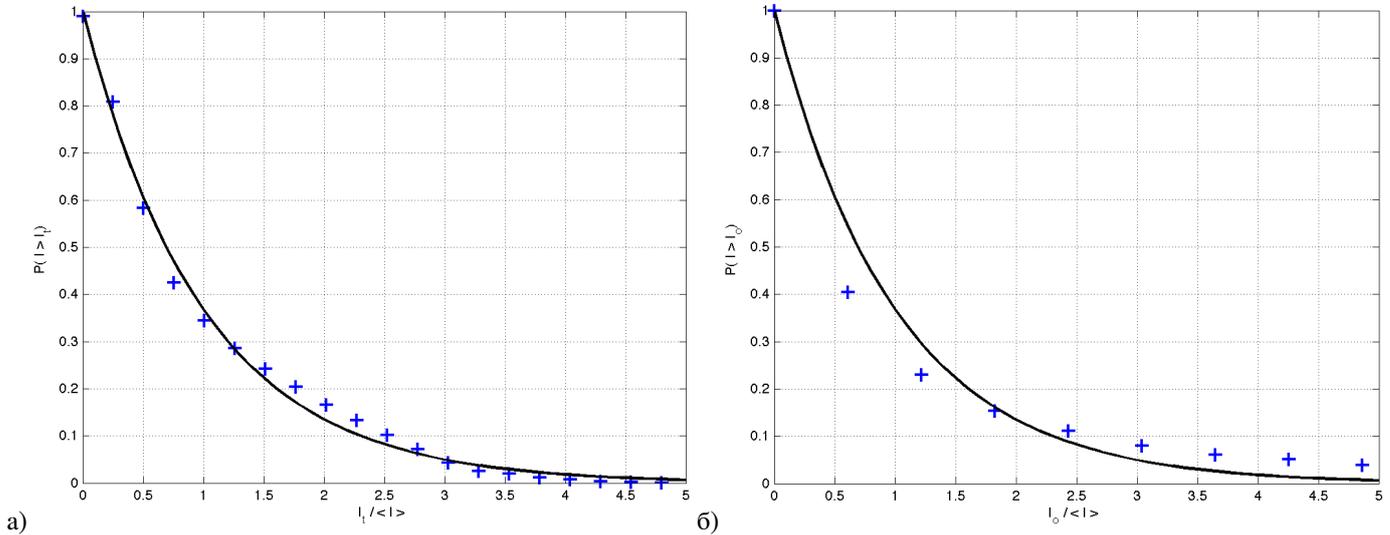

Рис. 9. Распределения вероятности значений интенсивности фокального пятна (в пределах контура уровня 1/2 максимума) по данным обработки изображений (крестики), а) измеренного изображения, б) расчетного изображения. Сплошной линией показана теоретическая кривая.
По вертикали - вероятность того что интенсивность $I$ имеет значение выше порогового значения $I_t$. По горизонтали - отношение порогового значения $I_t$ к средней величине интенсивности $<I>$.

Считается, что спекл-картина следует следующему распределению вероятностей [17]

$$P(I>I_t) = \exp^{-I/\langle I \rangle}, I \geq 0.$$

Эта зависимость получена в приближении случайного блуждания в комплексной плоскости и использует следующие допущения: во-первых, амплитуды и фазы слагаемых интерференционной картины являются статистически независимыми, во-вторых, фазы отдельных слагаемых равномерно распределены в интервале (-π, π), и, в-третьих, число слагаемых $N$ велико.

В данном случае число $N$=49, равно числу сегментов. Среднеквадратичное значение отклонение отраженного волнового фронта от сферы превышает длину волны, так что значения фаз разбросаны по всему интервалу (-π, π). Различие между рис. а) и б) в том, что для б) не выполнено первое условие. В случае а) независимая юстировка сегментов уменьшила согласованность фаз лучей, отраженных сегментов. При юстировке в видимом свете фаза не принималась во внимание, а лишь достигалось совпадение отраженных пятен света. Как результат, кривая а) гораздо лучше соответствует теоретической кривой, чем б).

## 8. Выводы

Полученные на двух длинах волн двумерные распределения отраженного от зеркала сигнала удовлетворительно совпадают с проделанным расчетом, включая отдельные области концентрации излучения. Наличие некоторой ступенчатости в профилях огибающих по видимому может быть связано как с недостаточной точностью/согласованностью угловых положений отдельных сегментов по вращению вокруг горизонтальных (см. рис. 1) осей лежащих в плоскости падения, так и с присутствием лавсановой пленки на трех из семи сегментов. (Пленка была оставлена для оптического контроля правильности ориентации формирователя как целого).

Сам этот факт, как и достаточно резкая неоднородность плотности мощности в области фокусировки не является помехой для интерпретации результатов измерений нагрева частиц в мощном терагерцовом пучке. В этом случае интегральное излучение усреднено не только вдоль измерительных хорд, поскольку тепловые постоянные доступных для измерения в настоящий момент пылинок [4, 11] сравнимы со временем их смещения на величину порядка расстояния между максимумами поля. Таким образом, мы получим хорошее усреднение по всему объему перетяжки. Разработанный формирователь вполне технологичен и его изготовление не является уникальной по сложности задачей.

Авторы полагают, что такие формирователи могут быть с успехом применены и во многих других случаях использования мощного ЛСЭ терагерцового диапазона Новосибирского центра фотохимических исследований.

# Литература

1. Каценеленбаум Б.З. Квазиоптические методы формирования и передачи миллиметровых волн. УФН, т.83, с. 81-105, 1964.

2. Oliver Lodge, Signalling Across Space Without Wires. Fleet Street, London, U.K.: "The Electrician" Printing & Publishing Company, 4th Ed., p. 83, 1908.

3. Винокуров Н.А. Состояние дел и перспективы лазера на свободных электронах Сибирского центра фотохимических исследований // Первое рабочее совещание «Генерация и применение терагерцового излучения»: Сборник трудов, 24 - 25 ноября 2005 г., ИЯФ им. Г.И. Будкера СО РАН, ИХКиГ СО РАН; Отв. ред. Б.А.Князев. - Новосибирск, с. 5-10, 2006.

4. Г.Д.Богомолов, В.Д.Громов. Экспериментальные исследования прохождения терагерцового излучения через взвешенную пыль. Постановка задачи и оценки. «Генерация и применение терагерцового излучения», 24-25 ноября 2005 г., ИЯФ им. Г.И. Будкера СО РАН, ИХКиГ СО РАН; Сборник трудов первого рабочего совещания. Отв. ред. Б.А.Князев. - Новосибирск, с. 85-95, 2006.

5. Meny, C.; Gromov, V.; Boudet, N.; Bernard, J.-Ph.; Paradis, D.; Nayral, C. Far-infrared to millimeter astrophysical dust emission. I. A model based on physical properties of amorphous solids. Astronomy and Astrophysics, v.468, pp.171-188, 2007.

6. Gold, B., Bennett, C. L., Hill, R. S. et al. Five-Year Wilkinson Microwave Anisotropy Probe Observations: Galactic Foreground Emission. Astrophys. J. Suppl., v. 180, pp. 265-282, 2009.

7. Usachev, A. D.; Zobnin, A. V.; Petrov, O. F.; Fortov, V. E. et al. Formation of a Boundary-Free Dust Cluster in a Low-Pressure Gas-Discharge Plasma. Phys. Rev. Lett., v. 102, 045001, 2009.

8. Wang, Zhehui; Mansfield, D.; Roquemore, L. et al. Applications and Progress of Dust Injection to Fusion Energy. MULTIFACETS OF DUSTRY PLASMAS: Fifth International Conference on the Physics of Dusty Plasmas. AIP Conference Proceedings, v. 1041, pp. 135-138, 2008.

9. Henning, T.; Mutschke, H. Optical properties of cosmic dust analogs: A review. Journal of Nanophotonics, 4(1), 041580, 2010.

10. J. Nelson. Segmented mirror telescopes. In: Optics in Astrophysics, ed. R. Foy and F. Foy. P. 61-72, Springer Netherlands, 2005.

11. Г. Д. Богомолов. Проектирование, изготовление и монтаж канала вывода терагерцового излучения лазера на свободных электронах. Отчет по программе РАН «Электромагнитные волны терагерцового диапазона». с. 87-92, 2005.

12. Б. А. Князев, В. С. Черкасский. Отражающие дифракционные оптические элементы и их применение для управления излучением терагерцового лазера на свободных электронах. Вестник НГУ. Серия: Физика. Том 1, выпуск 2, с. 3-20, 2006.

13. Vacuum Ultraviolet Spectroscopy. James A.R. Samson, p. 9, Wiley, New York. 1967.

14. М. Борн, Э. Вольф, Основы оптики. М., Наука, 1973.

15. В.В.Завьялов, Г.Д.Богомолов. ПТЭ. Вып 3, с. 174. 1982.

16. В.В.Завьялов, Е. М. Зотова Е И. Шампаров. ПТЭ. Вып 3, с. 109-114, 2008.

17. Goodman J.W. Some Fundamental Properties of Speckle. J.Opt. Soc. Am., v. 66, pp. 1145-1150, 1976.

18. Вайнштейн Л. Электромагнитные волны. М. 1988.

---

Примечания

*) Это название диапазона все чаще используется вместо менее ясно определенных терминов - дальний инфракрасный и субмиллиметровый диапазоны.

**) Термин квазиоптика [1] относится к таким устройствам формирования электромагнитного излучения, в которых используются методы геометрической оптики - отражение лучей, фокусировка, но при этом решающую роль играют волновые эффекты - дифракция, интерференция. Сам термин (quasi-optical) был введен Оливером Лоджем [2] вскоре после начала исследования электромагнитных волн Герцем, а затем Дж. Бозе и А. С. Поповым.